\documentclass[prb,aps,showpacs,amsmath,amssymb,superscriptaddress,twocolumn,longbibliography]{revtex4-1}\usepackage{color}
\usepackage{bm}
\usepackage{graphicx}
\usepackage{braket}
\usepackage{multirow}
\usepackage[plainpages=false,pdfpagelabels,colorlinks=true,linkcolor=red,urlcolor=blue,citecolor=blue,pdftitle={Title},pdfauthor={},pdfdisplaydoctitle=true,pdfduplex=DuplexFlipLongEdge]{hyperref}

\newcommand\ba{\begin{eqnarray}}
\newcommand\ea{\end{eqnarray}}
\newcommand\be{\begin{equation}}
\newcommand\ee{\end{equation}}

\begin{document}
\title{Engineering skin effect across a junction of Hermitian and non-Hermitian lattice}
\author{Ranjan Modak}
\email{ranjan@iittp.ac.in} 
\affiliation{ Department of Physics, Indian Institute of Technology Tirupati, Tirupati, India~517619}
\begin{abstract}
 We study a system where the two edges of a non-Hermitian lattice with asymmetric nearest-neighbor hopping are connected with two Hermitian lattices with symmetric nearest-neighbor hopping.  
 In the absence of those Hermitian lattices, the majority of the eigenstates of the system will be localized at the edges, the phenomena known as the non-Hermitian skin effect. We show that once we connect it with the Hermitian lattices, for open boundary conditions (OBC), the localized states exist at the junction of the non-Hermitian and Hermitian lattice; moreover, the spectrum shows mobility edges that separate delocalized and localized states. On the contrary, mobility edges vanish for periodic boundary conditions (PBC), and the delocalized phase turns into a scale-invariant localized phase, where the localized states are still peaked at the junctions.  We also find that if the connected Hermitian lattices are thermodynamically large, in OBC, most of the states become delocalized, while in PBC, the system still shows the scale-invariant localized phase.   

\end{abstract}

\maketitle

\section{Introduction}

Non-Hermitian systems have been of great interest in physics in recent years, uncovering a wide
range of phenomena~\cite{nh_1,nh_2,nh_3,nh_4,pal2022dna,modak2023non}. This surge started with the inception of PT-symmetric non-Hermitian quantum mechanics more than two decades ago. While in general non-Hermitian Hamiltonians have complex eigenvalues, by replacing the Hermiticity condition on the Hamiltonian with a much less constraining condition of space-time reflection symmetry known as PT-symmetry~\cite{Bender_2002, doi:10.1080/00107500072632, KHARE200053}, it was shown that the spectrum can be completely real. Such systems described by PT-invariant non-Hermitian Hamiltonians can typically be divided into two categories, one in which the  eigenvectors respect PT symmetry and the entire spectrum is real, and the other in which  the whole spectrum or a part of it is complex and the eigenvectors do not respect the PT symmetry.  These are known as PT-unbroken and broken phases, respectively. The  phase transition point is known as the exceptional point (EP). Recently, it has been demonstrated that there exists a Hermitian counterpart for the non-Hermitian systems in the PT-symmetric phase by defining a suitable inner-product~\cite{nogo,un_23}, 
while the PT-broken phase can't be mapped into any Hermitian system.  
This field of PT-symmetric non-Hermitian physics received a huge boost when the consequence of PT transition was observed experimentally in various systems~\cite{exp_pt_1,exp_pt_2,exp_pt_3,exp_pt_4,exp_pt_5,exp_pt_6}.

Moreover, in the last few years, significant importance has been given to
the understanding of the topological phases of non-Hermitian systems~\cite{RevModPhys.93.015005,PhysRevX.8.031079,PhysRevX.9.041015}, which shows a breakdown of
the correspondence between bulk topology and the existence of boundary modes~\cite{PhysRevLett.116.133903}. Subsequently, theoretical studies were successful in formulating a generalized
bulk-boundary correspondence  which is also consistent with experiments~\cite{nht1,nht2,nht3,nht4,nht5,nht6,nht7}.
In this context, the non-Hermitian skin effect (NHSE) has also brought lots of attention to the community. 
It has been shown that for certain non-Hermitian systems,  the majority of eigenstates of a non-Hermitian Hamiltonian are localized at boundaries~\cite{nht5,skin1,skin2,skin3}. The interplay of the NHSE with topology is something that makes this effect particularly interesting ~\cite{skin_topo}, and has been observed in experiments as well~\cite{zou2021observation}. These localized edge states in non-Hermitian systems have a topological origin that makes  them different from Anderson localized states in disordered systems~\cite{al2, al1}.  Moreover, understanding the interplay between NHSE and disordered potential also has been an active area of research~\cite{molignini2023anomalous, PhysRevA.95.062118}.
All these notions are not only theoretically enriching; there are many
potential technological applications of the non-Hermitian skin effect. Recently, a topological switch driven by  
the non-Hermitian skin effect has been proposed in
cold-atom systems~\cite{switch}, which is controlled by atom loss. Due to the extreme sensitivity of the boundary condition, NHSE can also lead to applications in quantum sensors~\cite{qs1,qs2}.

\begin{figure}
    \centering
    \includegraphics[width=0.46\textwidth]{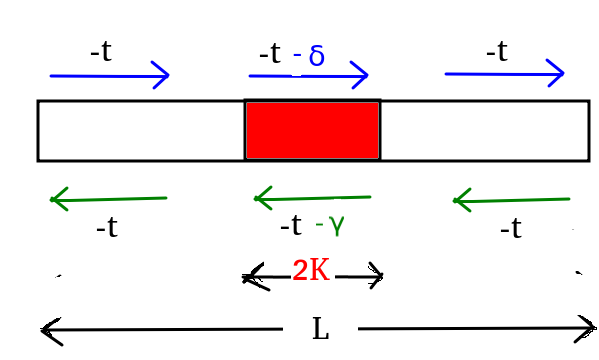} 
     
    \caption{Schematic diagram of our system (Hamiltonian Eqn.~\eqref{hamiltonian}). $-t$, $-t-\delta$  and $-t-\gamma$ indicate nearest-neighbour hopping strength. For $\delta\neq \gamma$, the system becomes non-Hermitian.}
    \label{fig0}
\end{figure}
\begin{center}
\begin{table}
\begin{tabular}{|c|| c |c |} 
 \hline
   &  OBC & PBC \\ [1.5ex] 
 \hline\hline
 $|Re[E]|<E_c$ & Delocalized phase &   \multirow{5}{8.0em}{Scale invariant localized phase; $\langle n_j\rangle \sim e^{-|j-(L/2-k)|/\xi(L)}$; $\xi(L)\propto L$}\\ [15ex]
 \hline
 $|Re[E]|>E_c$  &  \multirow{2}{8.0em}{Localized phase; $\langle n_j\rangle \sim e^{-|j-(L/2-k)|/\xi}$} & \multirow{2}{8em}{Localized phase; $\langle n_j\rangle \sim e^{-|j-(L/2-k)|/\xi}$} \\ [11ex]
 \hline
  \hline
  \end{tabular}
\caption{Table shows the energy-dependent phase diagram of the eigenstates of the Hamiltonian ~\eqref{hamiltonian} for both open boundary condition (OBC) and periodic boundary condition (PBC).}
\label{table1}
\end{table}
\end{center}

However, for all practical application purposes, one needs to find out a fine-tuned version of the skin effect.
Usually, the skin effect manifests localized states at the edges. 
Our main goal here is to engineer a system that gives control to the position of the localized state. Hence, we study a system where a non-Hermitian lattice is connected with the Hermitian lattices on both sides (see Fig.~\ref{fig0} for a schematic diagram); we find that at the junction of the Hermitian and non-Hermitian lattice, one observes the localized states. Hence, by changing the location of the non-Hermitian lattice, one can control the position of such localized states. The systems also seem extremely sensitive to boundary conditions, as most non-hermitian systems are. We outline our main results 
in Table.~\ref{table1}. We find for open boundary conditions (OBC), 
the systems support mobility edges~\cite{me1,me2,me3,me4} (coexistence of localized and delocalized states in the spectrum), and for periodic boundary conditions (PBC), the delocalized phase vanishes, and it turns into scale-invariant localized phase. However, remarkably, all the localized states peak at the junction.

The manuscript is organized as follows. We define our system in Sec. II. We elaborate on our main results in Sec. III. Finally, we discuss the conclusion and future prospect of our study in Sec. IV. 

\begin{figure}
    \centering
    \includegraphics[width=0.46\textwidth]{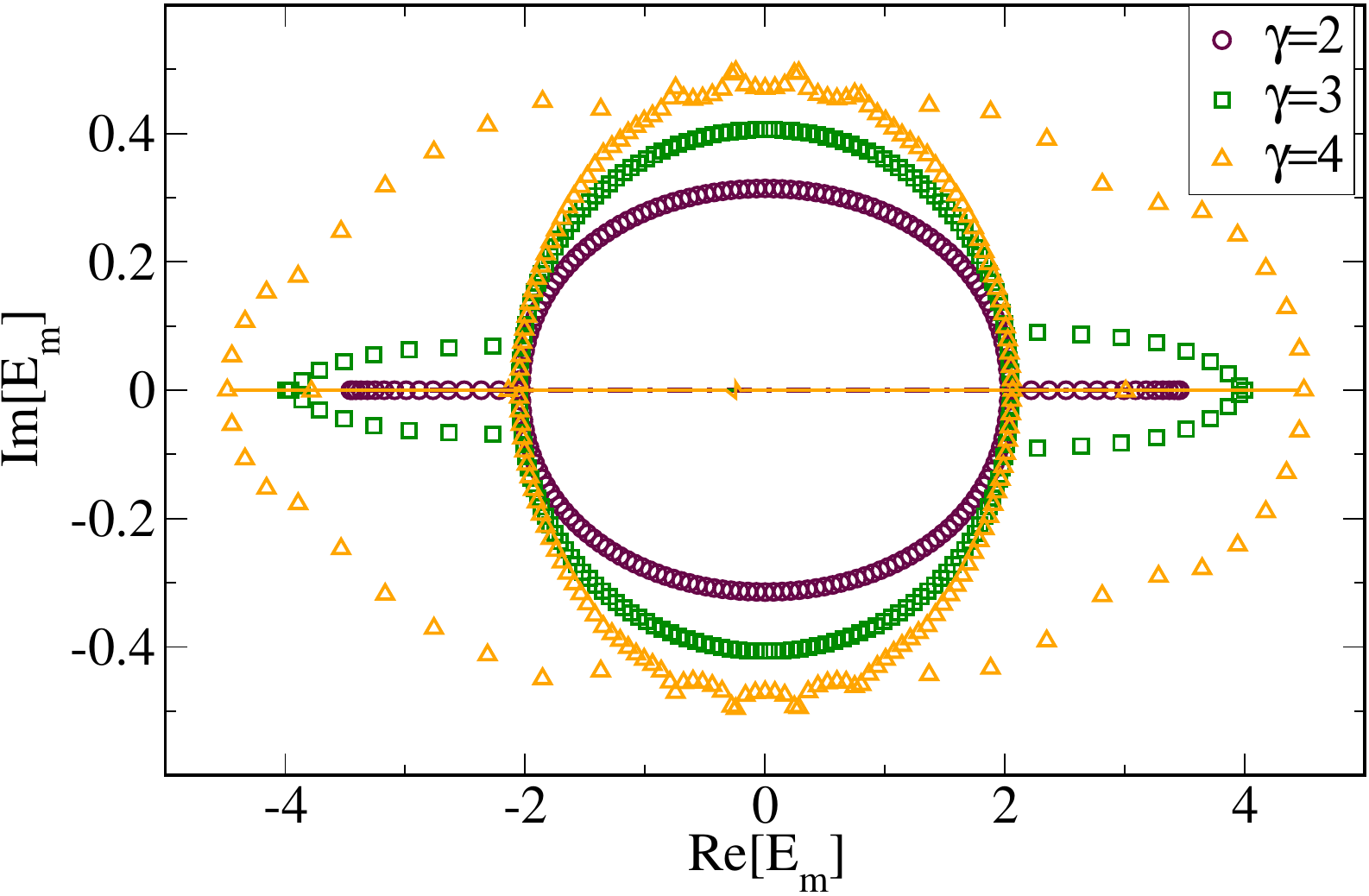} 
     
    \caption{Imaginary part of energy eigenvalues vs. real part of the eigenvalues for different values of $\gamma$ and $L=200$. Symbols correspond to PBC, and lines correspond to OBC.   }
    \label{fig1}
\end{figure}
\section{Model}
We study a system where edges of the non-Hermitian lattice of size $2k$ with asymmetric 
 nearest-neighbor hopping is connected with two Hermitian lattices of sizes $L/2-k$ each with symmetric nearest-neighbor hopping (see the schematic in Fig.~\ref{fig0}).
 The Hamiltonian (which is also non-Hermitian as a whole) is given by, 
 \begin{eqnarray}
 {H}&=&H_1+H_{nh}+H_2,~~ \text{where} \nonumber \\
 H_1&=&-\sum _{j=1}^{L/2-k-1}t(\hat{c}^{\dag}_j\hat{c}^{}_{j+1}+\text{H.c.})  \nonumber \\
 H_{nh}&=&-\sum _{j=L/2-k}^{L/2+k}(t+\gamma)\hat{c}^{\dag}_j\hat{c}^{}_{j+1}+(t+\delta)\hat{c}^{\dag}_{j+1}\hat{c}^{}_{j} \\
 H_2&=&-\sum _{j=L/2+k}^{L-1}t(\hat{c}^{\dag}_j\hat{c}^{}_{j+1}+\text{H.c.}), \nonumber 
 \label{hamiltonian}
\end{eqnarray}
where, $\hat{c}^{\dag}_j$ ($\hat{c}_{j}$) is the fermionic creation (annihilation) operator at site $j$, $\hat{n}_j=\hat{c}^{\dag}_j\hat{c}_{j}$ is the number operator. We introduce asymmetric  hopping by choosing $\gamma\neq \delta$. 
 In the results section, we report the results for both open boundary conditions (OBC) and periodic boundary conditions (PBC). Note that we add a Hermitian boundary term $-t(\hat{c}^{\dag}_L\hat{c}^{}_{1}+\hat{c}^{\dag}_1\hat{c}^{}_{L})$
with $H$ to implement PBC. 
The Hamiltonian ~\eqref{hamiltonian}  under PBC is exactly solvable in the Hermitian limit, i.e., $\gamma=0$ and $\delta=0$. The Hamiltonian reads as,  
\begin{eqnarray}
H(\gamma=0,\delta=0)=-\sum_{i}t(\hat{c}_{i}^{\dag}\hat{c}_{i+1}+\text{H.c}).
\label{tb}
\end{eqnarray}
  The Hamiltonian ~\eqref{tb} can be diagonalized in the momentum 
basis and  can be written as, 
\begin{eqnarray}
H(\gamma=0,\delta=0)=\sum_{k} \epsilon_k \hat{c}_{k}^{\dag}\hat{c}_{k}
\label{tb1}
\end{eqnarray}
where, $\epsilon _k=-2t\cos k$. $\hat{c}_{k}^{\dag}$ and $\hat{c}_{k}$ are fermionic creation and annihilation operators in momentum basis. Also, in the limit $t=0$, the Hamiltonian shows a non-Hermitian skin effect; as a result, all states become 
localized at the edge~\cite{zhang2022review,lin2023topological}. We choose $L$ as an even number, $\gamma >0$, $\delta=0$, and  $t=1$ for all the calculations in the main text. Next, we discuss the properties of the eigenstates of the Hamiltonian ~\eqref{hamiltonian} in the following section.

\begin{figure}
    \centering
    \includegraphics[width=0.46\textwidth]{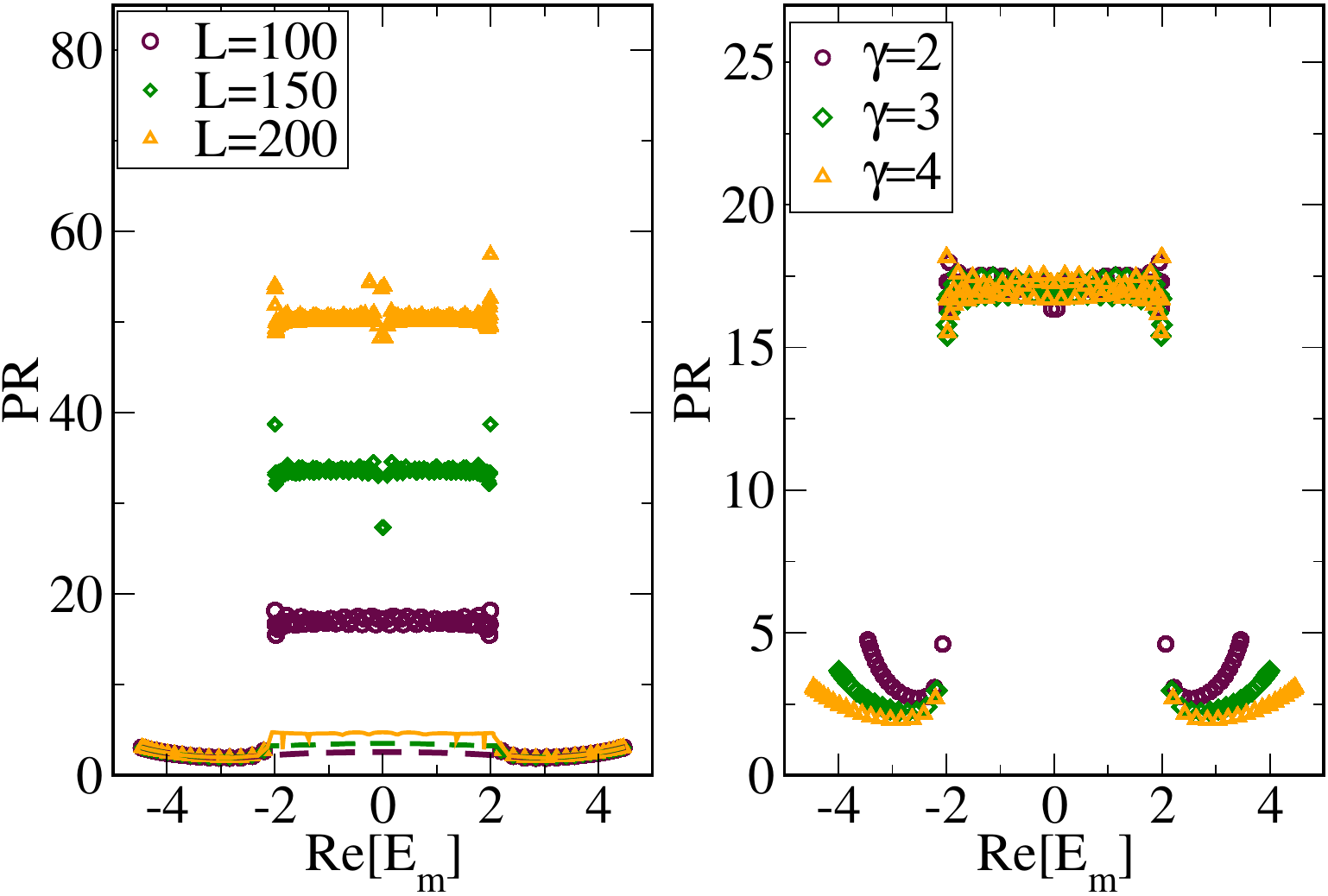} 
     
    \caption{(Left panel) Shows the variation of participation ratio (PR) with the real part of the energy for different values of $L$ and fixed $\gamma=4$. Symbols are for OBC, and lines are for PBC. (Right panel) Shows the variation  of PR for OBC for different values of $\gamma$ and for fixed $L=200$. We fix $k=25$.}
    \label{fig2}
\end{figure}

\begin{figure}
    \centering
    \includegraphics[width=0.46\textwidth]{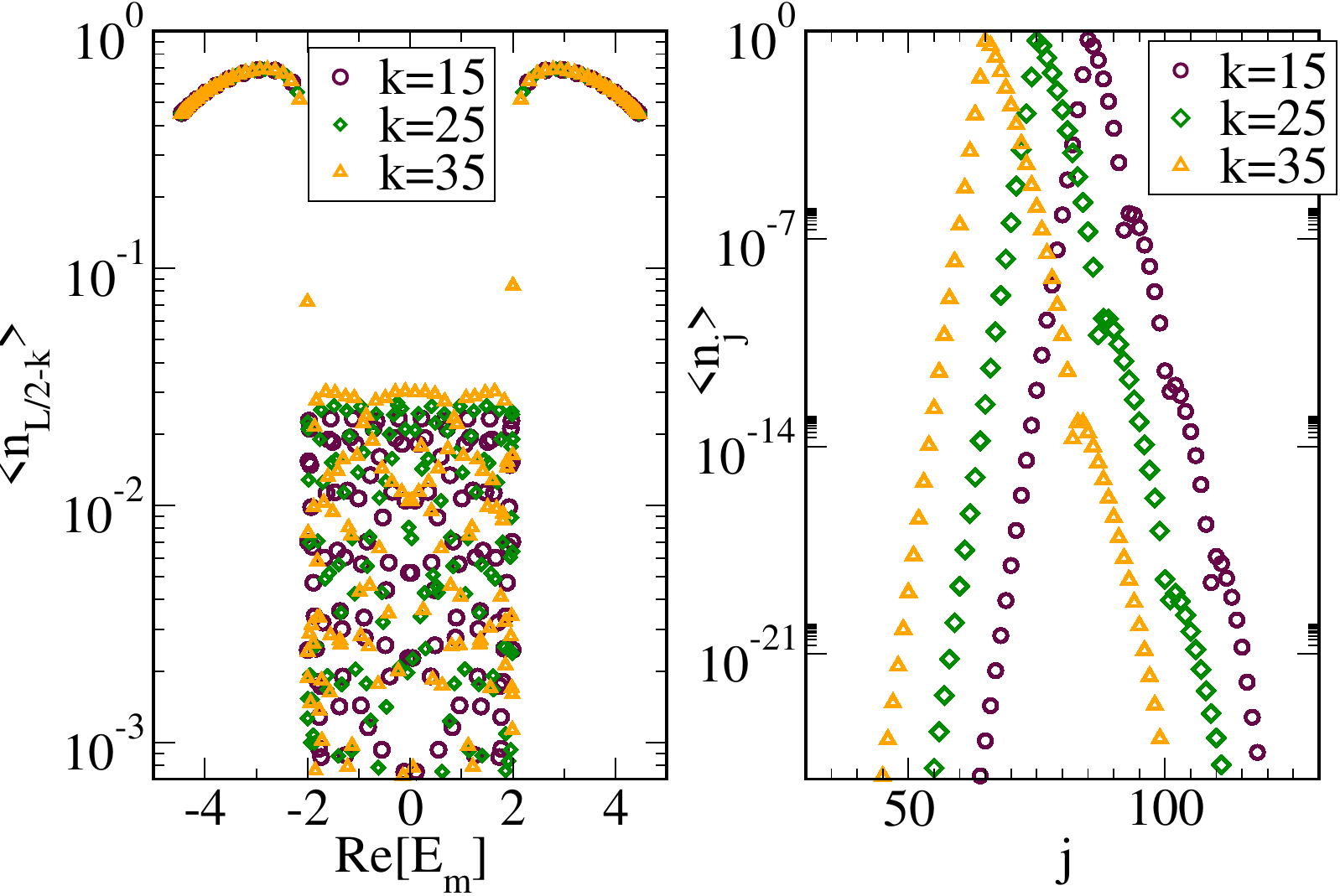} 
     
    \caption{(Left panel) shows the variation of $\langle n_{L/2-k}\rangle$ with the real part of energy eigenvalues for different values of $k$.
    (Right panel) shows the variation of $\langle n_j\rangle$ with $n_j$ for $4^{th}$ eigenstates for different values of $k$. We fix $L=200$, $\gamma=4$,  and use OBC.}
    \label{fig3}
\end{figure}

\begin{figure}
    \centering
    \includegraphics[width=0.46\textwidth]{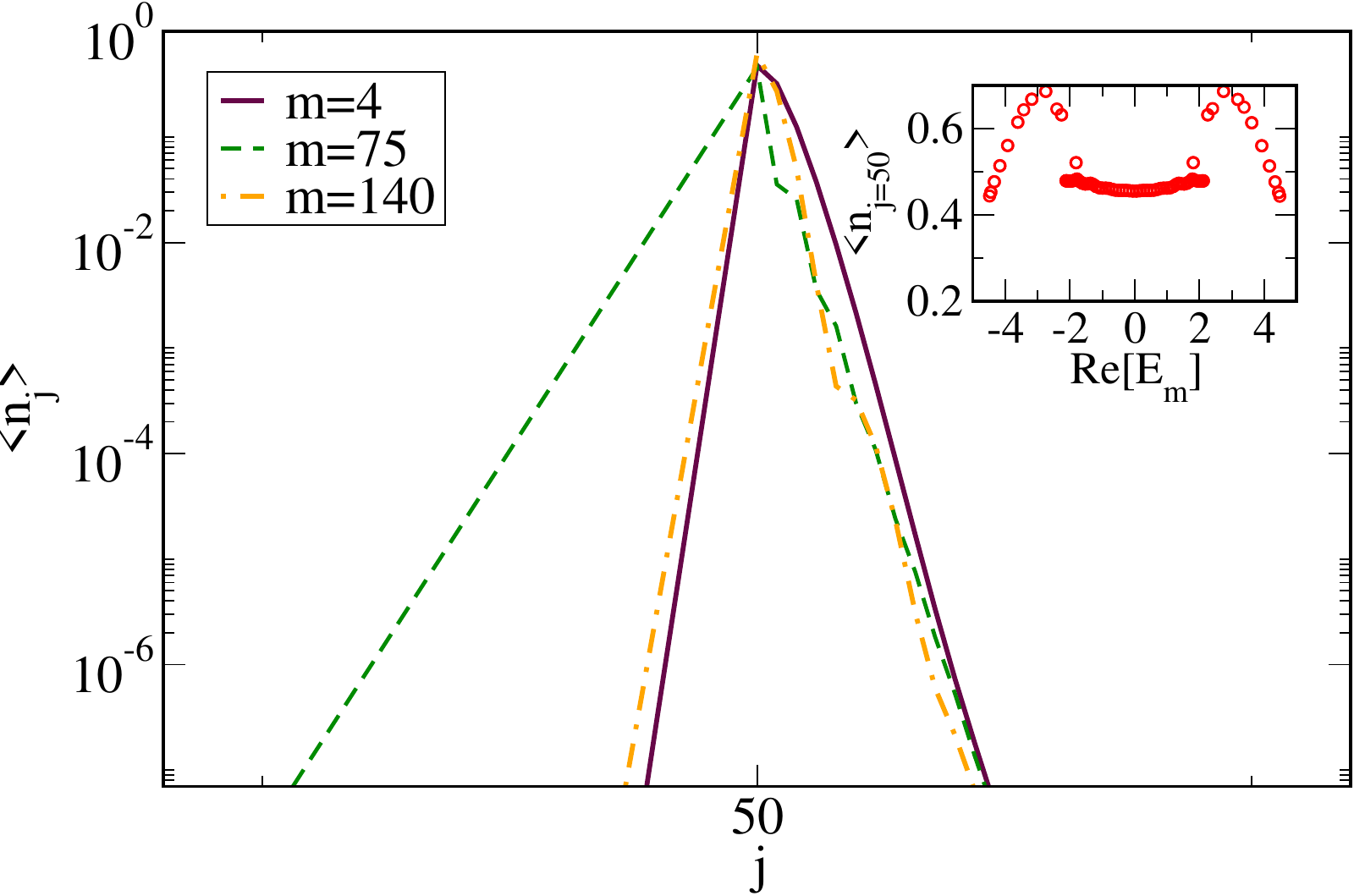} 
     
    \caption{Variation of $\langle n_j\rangle$ with site $j$ for $L=150$, $k=25$, $\gamma=4$, and  $m^{th}$ right-eigenstates. (Note eigenstates are arranged in ascending order of real part of energy eigenvalues) Inset shows the variation of $\langle n_{j=50}\rangle$ with the real part energy eigenvalues. These results are for PBC.  }
    \label{fig4}
\end{figure}

\begin{figure}
    \centering
    \includegraphics[width=0.46\textwidth]{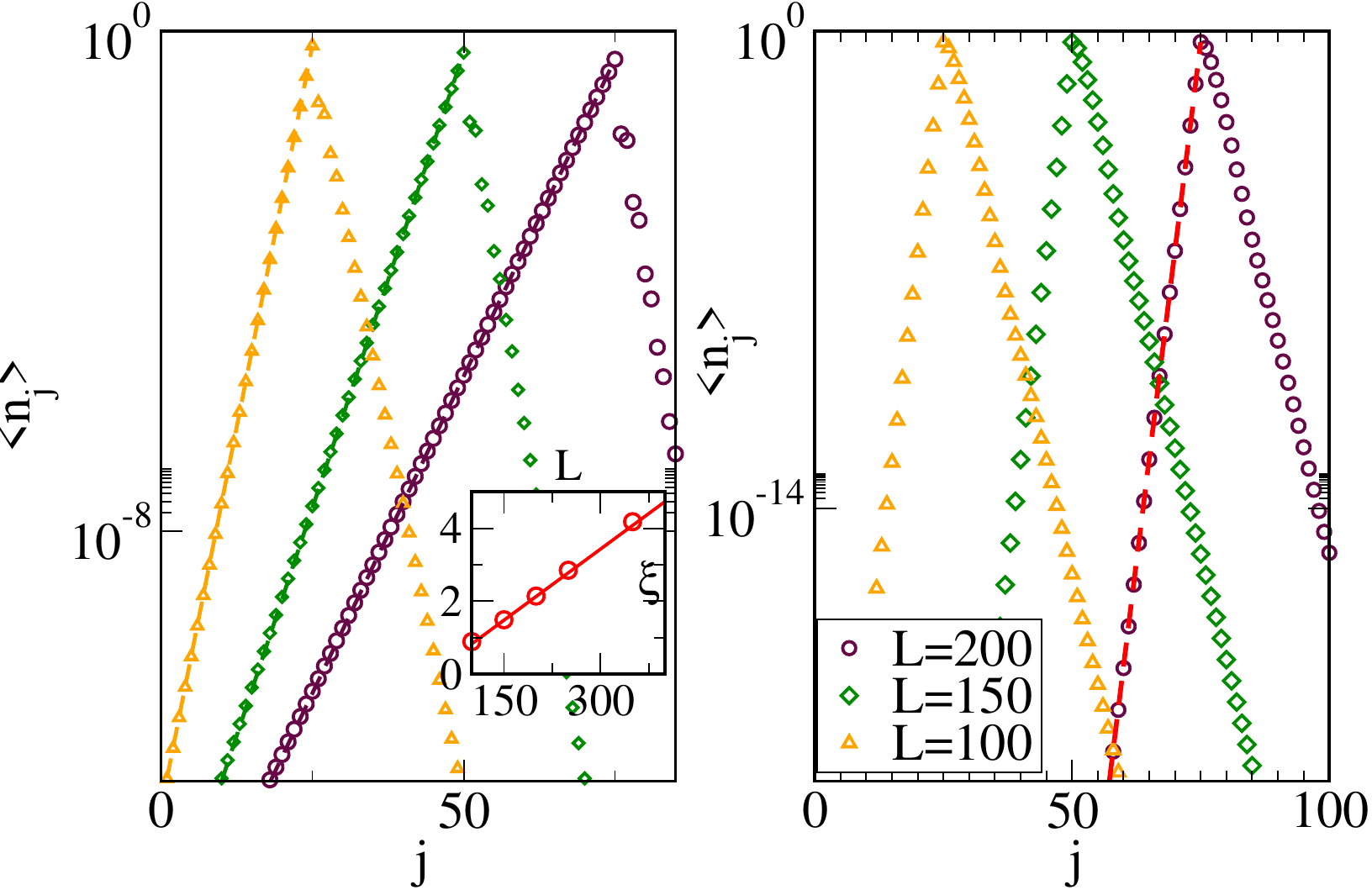} 
     
    \caption{Variation of $\langle n_j\rangle$ with site $j$ for different values of $L$, $k=25$, $\gamma=4$. 
    Results are for states $|Re[E]|<E_c$ (left panel) and $|Re[E]|>E_c$
    (right panel). 
    Dashed lines are exponential fitting curves i.e. $\langle n_j \rangle \sim \exp(-|x-(L/2-k)|/\xi)$.
    Inset shows linear scaling of localization length $\xi$ with $L$ for states $|Re[E]|<E_c$. These results are for PBC.  }
    \label{fig5}
\end{figure}
\section{Results}
First, we focus on the energy spectrum of the Hamiltonian ~\eqref{hamiltonian}. Note that our Hamiltonian is non-Hermitian; hence eigenvalues $E_m$ in principle can be complex numbers as well. We show in Fig.~\ref{fig1} the variation of the imaginary parts of the eigenvalues $E_m$ with its own real parts for both OBC and PBC. We find the eigenvalues are almost real for OBC; however, as soon as we introduce PBC, they become complex and form a loop in the complex plane of $E_m$. Usually, non-Hermitian systems are extremely sensitive to boundary conditions, and a similar type of eigenvalue spectrum  has also been observed for non-Hermitian models that manifest skin effect~\cite{PhysRevX.8.031079}. 

Next, we focus on the properties of eigenstates. Note that our system as a whole is also non-Hermitian. Hence, its right eigenstates and left eigenstates are not the same. We define an energy-dependent  localization-delocalization measure using the right eigenvectors, commonly known as participation ratio (PR)~\cite{PhysRevA.101.063612,PhysRevB.90.060205},

\begin{eqnarray}
\text{PR}(E_m)=\frac{(\sum_j| c^m_j|^2)^2}{\sum_j| c^m_j|^4}, 
\label{pr}
\end{eqnarray}
where, $|\psi_m^{R}\rangle=\sum_j c^m_j |j\rangle$ the right eigenvector satisfies $H|\psi_m^{R}\rangle=E_m|\psi_m^{R}\rangle$ (where $|j\rangle$ represents the site basis). PR is expected to be small $O(1)$ (does not scale with $L$) in the localized phase and scales linearly with $L$ in the delocalized phase. Figure.~\ref{fig2} (left panel) shows that for OBC and $|Re[E]|<E_c=2$ (which is the half bandwidth of the Hermitian lattice), all states are delocalized, and the rest of the states are localized. Hence, the spectrum shows mobility edges at $E=\pm E_c$. Remarkably, as we introduce PBC, mobility edges disappear (PR for PBC is shown using lines in Fig.~\ref{fig2} left panel). In the right panel of Fig.~\ref{fig2}, we plot the data for different values of $\gamma$, proving the robustness of our results.  

While the value of PR and its scaling with $L$ tell us whether a state is localized or delocalized, it does not tell us the structure of the states. The question, like where the localized states are peaked, can't be addressed for PR. Hence, 
next, we investigate structures of localized states in the case of OBC. In the limit, $t=0$, one observes the non-Hermitian skin effect, and the eigenstates are localized at site $j=L/2-k$ (for $\gamma>\delta$) and $j=L/2+k$ (for $\delta>\gamma$, which we discuss in the Appendix). Here, we find similar evidence for the localized states (see Fig.~\ref{fig3}). The right panel of Fig.~\ref{fig3} clearly shows that for localized states under OBC $\langle n_j\rangle\sim \exp(-|j-(L/2-k)|/\xi)$.  On the other hand, in case of PBC,  for eigenstates $|Re[E]|>E_c$. once again $\langle n_j\rangle\sim \exp(-|j-(L/2-k)|/\xi)$ (see Fig.~\ref{fig4} the results for $m=4$ and $140$). However, for  eigenstates $|Re[E]|<E_c$,

\begin{eqnarray}
\langle n_j\rangle&\simeq& \exp(-(L/2-k-j)/\xi_<),~~ \text{for}~~ j<L/2-k \nonumber \\
&\simeq& \exp(-(j-L/2+k)/\xi_>),~~ \text{for}~~ j>L/2-k. \nonumber \\
\label{nj}
\end{eqnarray}
Figure.~\ref{fig5} confirms that $\xi$, $\xi_>$ don't depend on $L$, while $\xi_<$ scales linearly with $L$ (see inset of Fig.~\ref{fig5}). Hence, we identify the  $|Re[E]|<E_c$ phase (for PBC) as the scale-invariant localized phase. To strengthen our claim that all the localized states for both PBC and OBC are peak at the junction, i.e., at site $L/2-k$, we compute $\langle n_{L/2-k}\rangle=\langle\psi^{R}|\hat{n}_{L/2-k}|\psi^{R}\rangle/\langle\psi^{R}|\psi^{R}\rangle$ for all the states. Figure.\ref{fig3} (left panel) clearly shows that for the OBC, the states which are localized ( when $|Re[E]|> E_c$), $\langle n_{L/2-k}\rangle$ very close to $1$, while for delocalized states the value is much smaller (which is expected given the states are not peaked at the junction). The inset of Fig.~\ref{fig4} shows similar results for the PBC. Due to the absence of a true delocalized state (mobility edge is also absent) and all the states are localized and peaked at the junction, under PBC, $\langle n_{L/2-k}\rangle$ value is $O(1)$ for all the states. It confirms that the localized states peak at the junction of the non-Hermitian and Hermitian lattice. 

\textit{Thermodynamic limit:}
Next, we discuss what happens in the thermodynamic limit. Note that one can approach the thermodynamic limit for our system in the following two ways, 
1) taking $L, k\to\infty$ limit keeping $k/L$ finite, 2) taking only $L\to \infty$ limit but $k$ is finite, hence $k/L\to 0$. 
For the first scenario, the conclusions obtained from our finite-size results will remain true, i.e., for OBC, mobility edges exist (coexistence of both localized and delocalized states). For PBC, 
the delocalized phase turns into a scale-invariant localized phase. On the other hand, for the 2nd scenario, for OBC, the faction of localized states $f\to 0$. It implies that the number of localized states will be "measure zero" in the limit $L\to \infty$, effectively suppressing the effect of localized state in the system. 
The same thing will also happen for Hermitian systems (see Appendix for details). 
On the other hand, under PBC, the majority of the states will behave as scale-invariant localized states, and all the states will peak at the junction.  

\section{Conclusions}
In this work, we study the fate of the non-Hermitian skin effect if a non-Hermitian lattice with asymmetric hopping is connected with a Hermitian lattice on both sides. It is well established that without the Hermitian lattice connection, a non-Hermitian lattice of asymmetric hopping usually shows localized edge modes at the boundary, known as the non-Hermitian skin effect. We find here that under OBC, the localized states still exist at the junction of the non-Hermitian and Hermitian lattice. There also exist delocalized states separated by the so-called mobility edges in the spectrum. On the contrary, under PBC, mobility edges vanish, and the delocalized phase turns into a scale-invariant localized phase, i.e., localization length of the right wave-function scales with system size. Hence, we show by connecting the non-Hermitian lattice with the Hermitian one; it's possible to engineer the localized states 
at the junction of two such lattices, which gives us additional freedom to control the position of the localized states, which typically is only observed at the boundary for the non-Hermitian skin effect. 
In the thermodynamic limit (keeping non-Hermitian lattice size fixed), for OBC, the system will be completely delocalized; however, strikingly, for PBC, the system  will remain in a scale-invariant localized phase, where states will be peaked at the Hermitian and non-hermitian lattice junction.  We also like to point out that such a scale-invariant localized phase has also recently been observed in disordered non-Hermitian lattices, and it was referred to as the "Anomalous skin effect". 

Also, theoretically, it raises a broader question of the stability of the non-Hermitian skin effect of a finite-sized non-Hermitian lattice once connected with a delocalized Hermitian infinite bath. Unlike the Hermitian system, here, the fate of the non-Hermitian skin effect will depend on the boundary condition. For open boundary conditions, for a true infinite delocalized bath, the majority of states  of the whole non-Hermitian system will be delocalized, but in a ring geometry (PBC), the states will display scale-invariant localizations, and all  the states will be peaked at the junction between Hermitian and non-Hermitian lattice. In contrast, for the Hermitian system (assuming an Anderson-localized system is placed between two Hermitian delocalized lattices), independent of the boundary conditions, in the thermodynamic limit, most of the states will always be delocalized (see Appendix for the comparison with Hermitian systems).  
Further, It will be interesting to investigate the role of interactions and quasiperiodic potentials in such cases in future studies~\cite{hamazaki2019non, zhai2020many,wang2022non,suthar2022non}.

\section{Acknowledgements}
RM acknowledges the DST-Inspire fellowship by the Department of Science and Technology, Government of India, SERB start-up grant (SRG/2021/002152).
\appendix 
\section{Results for $\delta>0$ and $\gamma=0$}
\begin{figure}[h!]
    \centering
    \includegraphics[width=0.46\textwidth]{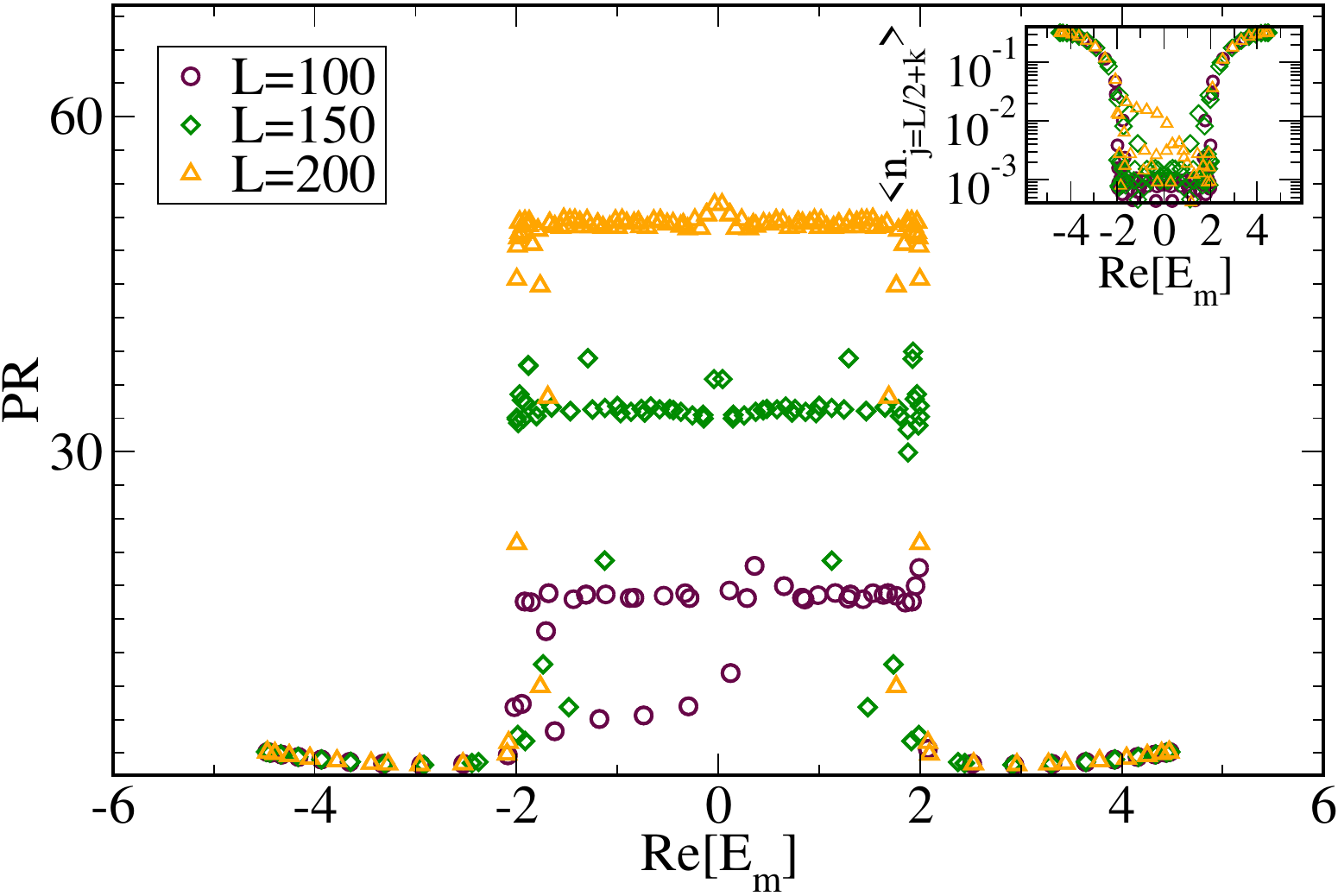} 
     
    \caption{Variation of PR with real part of the energy $E_{m}$
    for different values of $L$ and for 
    fixed $k=25$, $\delta=4$ and $\gamma=0$. 
    Inset shows the variation of $\langle n_{L/2+k}\rangle $
for different eigenstates. These results are for OBC.  }
    \label{fig6}
\end{figure}
\begin{figure}[h!]
    \centering
    \includegraphics[width=0.46\textwidth]{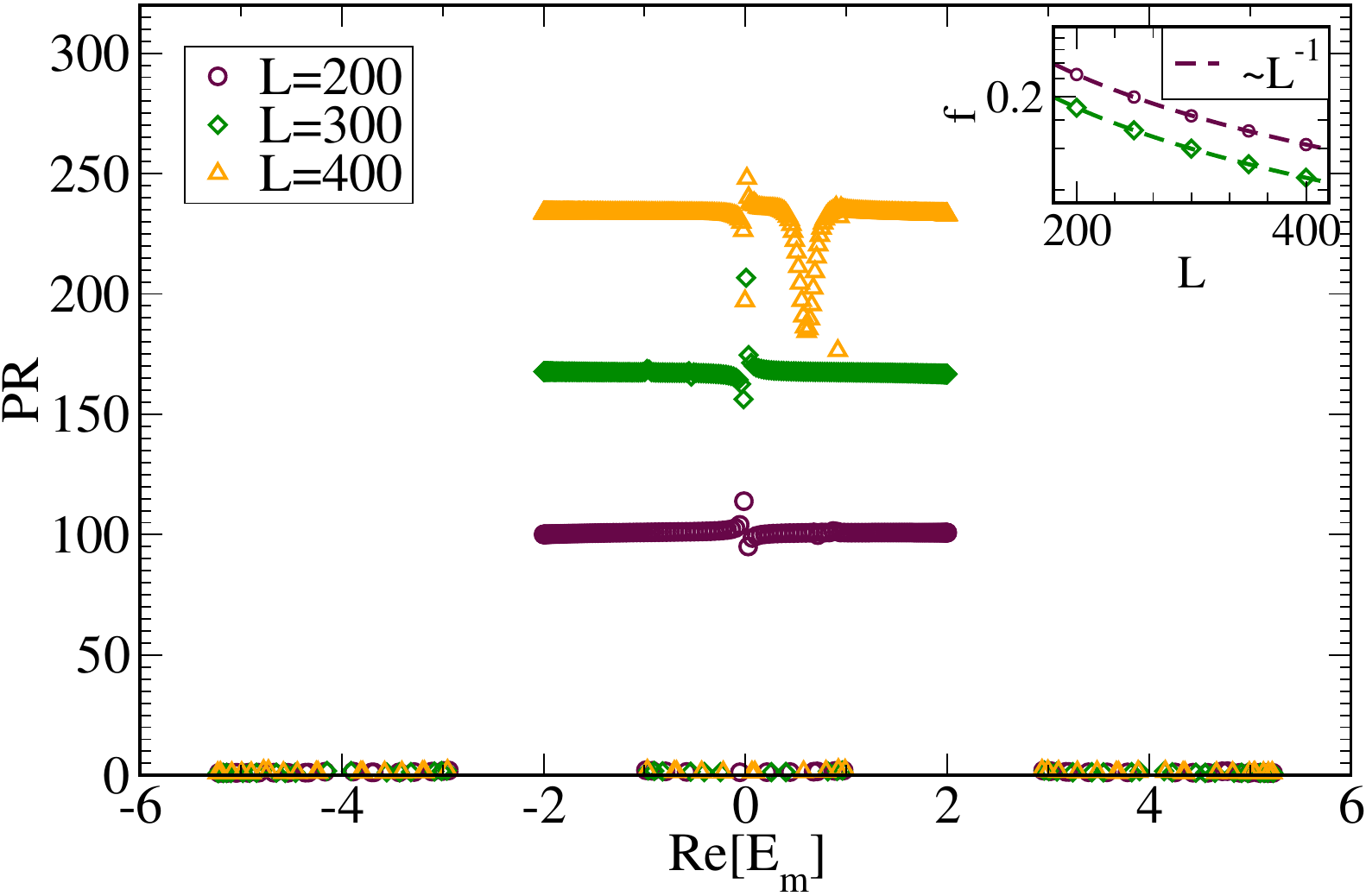} 
     
    \caption{Variation of PR with real part of the energy $E_{m}$ for different values of $L$ and for 
    fixed $k=25$, $V=5$ for the Hermitian Hamiltonian ~\eqref{hamiltonian_h} for PBC. 
    Inset shows the variation of  the fraction of localized states $f$ vs $L$ for the Hamiltonian $\tilde{H}$ (PBC) and the non-Hermitian  Hamiltonian ${H}$ (OBC) using circle and diamond symbols, respectively.}
    \label{fig7}
\end{figure}
While in the main text, we choose $\delta=0$ and $\gamma>0$, in this section, we demonstrate the results for  $\delta>0$ but $\gamma=0$. 
Figure.~\ref{fig6} clearly shows that similar to the main text results, the spectrum shows mobility edges for OBC. PR increases with system size for $|Re[E]|<E_c$ indicating delocalized states, while PR of the rest of the states do not scale with system size. In the inset, we plot the variation of $\langle n_{L/2+k}\rangle$ for different states. Results suggest that the localized states peaked at site $j=L/2+k$. While for the $\gamma>\delta=0$ case, the peaks of the localized states were at another junction i.e.  at site $j=L/2-k$ as shown in the main text.  \\

\section{Comparison with Hermitian systems}

We study a system that is described by the following Hermitian Hamiltonian, 
 \begin{eqnarray}
 \tilde{H}&=&\tilde{H}_1+H_{h}+\tilde{H}_2,~~ \text{where} \nonumber \\
 \tilde{H}_1&=&-\sum _{j=1}^{L/2-k-1}t(\hat{c}^{\dag}_j\hat{c}^{}_{j+1}+\text{H.c.})  \nonumber \\
 H_{h}&=&-\sum _{j=L/2-k}^{L/2+k}t(\hat{c}^{\dag}_j\hat{c}^{}_{j+1}+\hat{c}^{\dag}_{j+1}\hat{c}^{}_{j})+V\sum_{j=L/2-k}^{L/2+k}\cos(2\pi\beta j) \nonumber \\
 \tilde{H}_2&=&-\sum _{j=L/2+k}^{L-1}t(\hat{c}^{\dag}_j\hat{c}^{}_{j+1}+\text{H.c.}), \nonumber \\
 \label{hamiltonian_h}
\end{eqnarray}
where, $\hat{c}^{\dag}_j$ ($\hat{c}_{j}$) is the fermionic creation (annihilation) operator at site $j$, $\hat{n}_j=\hat{c}^{\dag}_j\hat{c}_{j}$ is the number operator, $\beta=(\sqrt{5}-1)/2$. For $V>2$ ($V<2$), all the states of $H_h$ are localized (delocalized)~\cite{aubry1980analyticity}. This model is famously known as the Aubry-Andre model, which in contrast to the true disorder, shows a delocalization-localization transition even in one dimension~\cite{aa1,aa2}. 
Figure.~\ref{fig7} shows the variation of PR for fixed $k=25$, $V=5$ but different values $L$ for PBC. 
Unlike our results in the main text, for Hermitian Hamiltonian, even in PBC, eigenstates in the middle of the spectrum remain delocalized. 
In the inset, we plot the fraction of localized state $f$ vs $L$. In the same inset, we also show the variation of $f$ for the non-Hermitian Hamiltonian ~\eqref{hamiltonian} for OBC, find that $f\to 0$ as $L\to \infty$ as $1/L$.  Note that to compute $f$, those states which don't scale with $L$ have been identified as localized states. Hence, our main results in the main text, i.e., the existence of scale-invariant localized phase under PBC while the absence of localization in OBC in the thermodynamic limit (keeping the size of the non-Hermitian lattice fixed) is purely a non-Hermitian quantum phenomena; it does not have any Hermitian counterpart. 

\bibliography{ref,ref_3}
\end{document}